\documentclass[12pt]{article}

\usepackage{rotating}

\usepackage{multirow}
\usepackage{amsmath,amssymb,graphicx, graphics, color,verbatim} 
\usepackage{cite}
\usepackage{latexsym} 
\usepackage{cancel}
\usepackage{hyperref}

\usepackage{graphicx}
 \usepackage{caption}
 \usepackage{subcaption}
 \usepackage{slashed}

\usepackage{enumerate} 

\usepackage[normalem]{ulem} 

\pdfoutput=1
\usepackage{epsf}
\usepackage{pstricks} 
\usepackage{comment}

\usepackage{array}

\newcommand{\centeron}[2]{{\setbox0=\hbox{#1}\setbox1=\hbox{#2}\ifdim
\wd1>\wd0\kern.5\wd1\kern-.5\wd0\fi \copy0
\kern-.5\wd0\kern-.5\wd1\copy1\ifdim\wd0>\wd1
                                   \kern.5\wd0\kern-.5\wd1\fi}}
\newcommand{\ltap}{\>\centeron{\raise.35ex\hbox{$<$}}
                           {\lower.65ex\hbox{$\sim$}}\>}
\newcommand{\gtap}{\>\centeron{\raise.35ex\hbox{$>$}}
                           {\lower.65ex\hbox{$\sim$}}\>}

\newcommand{\lsim}{\mathrel{\ltap}}

\newcommand\ZZ{\hbox{\zfont Z\kern-.4emZ}}
\font\zfont = cmss10 

\newcommand{\cref}[1]{Chapter \ref{c.#1}}

\newcommand{\ba}{\begin{array}}
\newcommand{\ea}{\end{array}}

\newcommand{\beq}{\begin{eqnarray}}
\newcommand{\eeq}{\end{eqnarray}}
\newcommand{\beqs}{\begin{eqnarray*}}
\newcommand{\eeqs}{\end{eqnarray*}}

\newcommand{\bal}{\begin{align}} 
\newcommand{\eal}{\end{align}}

\def\bi{\begin{itemize}}
\def\ei{\end{itemize}}
\def\ben{\begin{enumerate}}
\def\een{\end{enumerate}}
\def\bc{\begin{center}}
\def\ec{\end{center}}
\def\bt{\begin{table}}
\def\et{\end{table}}
\def\btb{\begin{tabular}}
\def\etb{\end{tabular}}

\def\co{{\mathcal O}}

\def\mass2{mass${}^2$}

  \newcommand{\ww}{
$W^+W^-$
}

\newcommand{\wz}{$W^\pm Z$~}

\newcommand{\zz}{$Z Z$~}

\newcommand{\pt}{$p_T$~}

\newcommand{\der}[1]{\mathrm{d} #1}
\def\ptv{ p_T^{\text{veto}} }
\def\MET{\mathrm{MET}}

\begin{document}
\bibliographystyle{unsrt}
\begin{titlepage}
\begin{flushright}
\small{YITP-SB-15-36}
\end{flushright}

\vskip2.5cm
\begin{center}
\vspace*{5mm}
{\huge \bf Precision diboson measurements and the interplay of \pt and jet-veto resummations}
\end{center}
\vskip0.2cm

\begin{center}
{Prerit Jaiswal$^{(1)}$, Patrick Meade$^{(2)}$, Harikrishnan Ramani$^{(2)}$}

\end{center}
\vskip 8pt

\begin{center}
{\it $^{(1)}$Department of Physics, Brown University\\ Providence, RI 02912\\}
{\it $^{(2)}$C. N. Yang Institute for Theoretical Physics\\ Stony Brook University, Stony Brook, NY 11794}\\
\vspace*{0.3cm}

\vspace*{0.1cm}

{\tt prerit\_jaiswal@brown.edu,meade@insti.physics.sunysb.edu, hramani@insti.physics.sunysb.edu}
\end{center}

\vglue 0.3truecm

\begin{abstract}

In this paper we demonstrate the agreement of jet-veto resummation and \pt resummation for explaining the \ww cross sections at Run 1 of the LHC, and in the future.  These two resummation techniques resum different logarithms, however via reweighting methods they can be compared for various differential or exclusive cross sections.  We find excellent agreement between the two resummation methods for predicting the zero-jet cross section, and propose a new reweighting method for jet-veto resummation that can be used to compare other differential distributions. We advocate a cross-channel comparison for the high-luminosity run of the LHC as both a test of QCD and new physics.

\end{abstract}

\end{titlepage}

\section{Introduction}
\label{s.intro} \setcounter{equation}{0} \setcounter{footnote}{0}
Run I of the LHC was an overwhelming success, the Higgs discovery\cite{Aad:2012tfa} completed the Standard Model (SM) of particle physics.  However, it also explicitly brought to the fore the question of naturalness in the SM.  In particular, the lack of any sign of physics beyond the SM (BSM) at Run 1 leaves a number of questions for  the prospects of discovery at Run 2 given the impressive exclusions at high masses.  Nevertheless, the LHC has also brought about a new opportunity for precision measurements at the electroweak (EW) scale and the opportunity to test the SM in ways that were essentially inaccessible before.  New precision measurements are crucial both for understanding the properties of the Higgs better, as well as for searching for new physics at the EW scale where the phase space of new physics would strongly overlap with the SM.

At Run 1 of the LHC the energy and luminosity were sufficient to start probing certain EW processes with unprecedented statistics, e.g. diboson production.  Probing diboson production is important for a number of reasons within the SM, as it is the main background for several of the most important Higgs search channels, and it can test the EW gauge structure of the SM.  Historically it is also useful for looking for deviations from the SM for instance in aTGCs(anomalous triple gauge couplings) and aQGCs(anomalous quartic gauge couplings) which can be related to a whole host of EW precision tests (EWPT).   Diboson production is also an important background for almost any model of new physics that has new EW charged particles or modifies and extends the EW gauge/Higgs structure of the SM in any way.  Given the ubiquitous importance of diboson production, it is necessary to improve both the theoretical and experimental understanding of the many channels within the SM.

In most diboson channels at run 1 and especially at run 2 there will be sufficient statistics such that all productions modes can be observed in leptonic final states making for relatively clean measurements.  In fact almost all measurements of the total inclusive cross section at run 1 agreed very well with the NLO QCD predictions.  However, there is far more information that can be gleaned from diboson channels than their overall rate alone.  Given that the production of a diboson pair is an uncolored final state, the QCD corrections to diboson production will have very similar predictions that roughly depend on the mass scale and the particles in the initial state production e.g. $q\bar{q}$ or $gg$.  For instance the transverse momentum distribution of the \ww, \zz and \wz channels all should be very similar and correlating between various channels can be a test of QCD.  This is similar to the program carried out at lower energies where Drell-Yan production and single $W^\pm$ productions can be correlated and predictions can be made that allow one to extract important EW measurements such as the $W^\pm$ mass.  Understanding diboson production in differential shape directions can test QCD, but it can also then be applied to searches/exclusions of new physics models as has been demonstrated in the \ww channel~\cite{Curtin:2013gta,Curtin:2014zua} and applying similar methods to the $t\bar{t}$ channel in~\cite{Czakon:2014fka}.  In this paper we begin to suggest a program of measurements and correlations amongst  EW diboson channels motivated by current higher order QCD calculations \cite{NNLOdiboson, Gehrmann:2014fva}. This program can be straightforwardly extended to processes beyond diboson production, but we focus on diboson production in this paper because of an anomaly that persisted from the Run 1 of the LHC.

At run 1 of the LHC almost all the diboson channels agreed with the SM NLO QCD predictions except for the \ww (ATLAS and CMS) and $W^\pm\gamma$(CMS only) and $W^\pm Z$(CMS only)  channels.  The \ww channel is particularly interesting because it consistently reported discrepancies with theoretical predictions both at 7 and 8 TeV, and both in ATLAS \cite{atlas7ww,atlas8ww} and CMS\cite{cms7ww,cms8ww}.  Importantly the excesses reported were not only in the overall rate but there were also shape discrepancies in many differential directions.  Many attempts to explain the excess  were put forward using both BSM physics\cite{Curtin:2012nn,Curtin:2013gta,Rolbiecki:2013fia,Curtin:2014zua,Kim:2014eva,Jaiswal:2013xra,Dermisek:2015vra, Dermisek:2015oja}, as well as higher order QCD corrections to the SM process.  The \ww channel is currently unique amongst the diboson channels as it employs a jet veto of $25$ GeV ($30$ GeV) for ATLAS (CMS) to reduce the $t\bar{t}$ background.  This implies that not only fixed order QCD corrections are important, but there can also be large logarithms that  need to be resummed as well.  The introduction of the jet veto is of course not the only reason that large logs may appear and need to be resummed, in certain differential directions it is crucial to include resummation to predict the shape accurately\footnote{There are also threshold logarithms associated with soft-gluon emissions. Threshold resummation and approximate NNLO results for \ww production were presented in \cite{Dawson:2013lya}.}.

Given the presence of the jet veto for the \ww channel it makes sense to perform jet-veto resummation to resum logs of the form $\ln \left(\ptv/M\right)$, where $M$ is the scale of the hard interaction, and see the effects on the \ww cross section.  This was carried out in~\cite{Jaiswal:2014yba} where it was shown that it does improve the agreement between the measured cross section and theoretical prediction.  However, using jet-veto resummation alone does not directly make predictions for other differential directions.  To describe other differential directions one must employ a reweighting of MC events, which we explore in this paper, or a joint resummation. These are both interesting and compelling avenues to pursue, because as stated earlier the experimental \ww measurements had shape discrepancies and not solely rate discrepancies.

One interesting differential direction reported by ATLAS was  $p_T$($\ell^+\ell^-+\MET$) which had a shape discrepancy particularly at low values of this variable where there were an excess of events~\footnote{CMS has not released a distribution of this,but it would be very useful if they did.} .  What makes the $p_T$($\ell^+\ell^-+\MET$) distribution particularly interesting is that it is essentially a proxy for the $p_T^{WW}$ given that it was measured in the fully leptonic channel.  It is well known to accurately predict the $p_T$ shape for EW final states at low $p_T$,  $p_T$ resummation must be used to go beyond fixed order calculations or MC parton shower predictions.  While naively $p_T$ resummation will not change the overall inclusive cross section at all~\cite{Bozzi:2005wk}, there is a strong correlation between the $\ptv$ and $p_T^{WW}$ when a jet veto is imposed.  For instance at NLO, the jet recoiling off the diboson pair has equal but opposite transverse momentum.  However, with $p_T$ resummation alone there are no jets, and hence this correlation can only be extended and observed by employing a reweighting procedure for instance as used in~\cite{D0:2013jba, deFlorian:2011xf}.  This was done for the \ww channel in~\cite{Meade:2014fca} and it was shown by reweighting with respect to the resummed $p_T^{WW}$, there were effects on the fiducial cross section which improved the agreement with the experimental data.  Subsequently, the full luminosity 8 TeV analysis by CMS\cite{Khachatryan:2015sga} employed the $p_T$-resummation curves reweighting from~\cite{Meade:2014fca} as well as the full NNLO cross-section\cite{Gehrmann:2014fva} and found good agreement between experiment and theory. Whereas the reweighting from NNLO is an overall normalization, the reweighting from $p_T$-resummation is a shape effect.

Because of the strong correlation between $\ptv$ and $p_T^{WW}$ it should be expected that the jet-veto resummation~\cite{Jaiswal:2014yba} and the $p_T$ resummation calculations~\cite{Meade:2014fca} should give similar results.  In~\cite{Banfi:2012yh} this was shown to be the case for Drell-Yan and Higgs production, with discrepancies occurring at higher orders because the correlation between $\ptv$ and $p_T^{WW}$ is weakened and depends on jet clustering effects.  Nevertheless a naive reading of \cite{Jaiswal:2014yba}  and \cite{Meade:2014fca} seemed to imply that there was a larger discrepancy between these methods than would be expected. In fact this led to further paper on QCD effects in these channels trying to explain the experimental discrepancy\cite{Monni:2014zra} and automate jet-veto resummation\cite{Becher:2014aya}.  Given the success of \cite{Khachatryan:2015sga} it is important to further study how well $p_T$ resummation captures the jet veto logs in the \ww process.  

In this paper we show that  \cite{Jaiswal:2014yba}  and \cite{Meade:2014fca} agree quite well when carefully compared using the same experimental variable~\footnote{The main naive discrepancy is due to how \cite{Jaiswal:2014yba} presented the effects of resumming additional $\pi^2$'s but there is no inherent discrepancy when making predictions for the fiducial cross section that ATLAS or CMS would measure.}.  We additionally investigate more generally the comparison between jet veto and $p_T$ resummation with the same scale choices and parameters to understand their correlation and interplay.   While this is useful for making predictions for jet-vetoed cross sections, it doesn't address other differential directions in particular why there are shape discrepancies in the fiducial cross section as well for \ww.  To investigate this we propose a new method to use Jet-Veto resummation to reweight MC samples to obtain a more accurate prediction of differential cross sections with a jet veto. We then study how well the predictions of this method compares with $p_T$ resummation.
The dependence of the agreement between these resummation reweighting methods for different jet radius is studied in detail for $R=0.4,0.5$ and for a large radius $R= 1$ where correlations are expected to be stronger. We also investigate the contributions from non-perturbative (NP) effects such as hadronization and Multi-Parton Interactions(MPI).

Based on this study the results can be extended to a better understanding of other diboson channels as well. For instance, while there has been extensive work on NP effects and scale choices for single vector boson processes, at the LHC this may now be carried out across even more channels. The detailed understanding of the \ww channel, which has high statistics but additional jet-veto complication, could then be used in conjunction with other diboson processes which are more rare but do not have a jet veto. Fitting across various channels at the high luminosity LHC could shed light on optimal resummation scale choices and modeling non perturbative factors as well as allowing for new opportunities to test QCD and search for new physics.  In particular the fact that at run 1 diboson channels other than \ww seemed to agree well with only NLO MC predictions, whereas \ww required NNLO+NNLL QCD calculations to be accurately described could provide a window into understanding how well the SM actually describes the data when theoretical predictions are {\em uniformly} applied.  The rest of the paper is organized as follows. In section 2, we briefly review Jet veto and $p_T$ resummation theory. In section 3, we introduce our new method to reweight using jet veto resummation and reweighting using $p_T$ resummation is reviewed. In section 4 we demonstrate the correlation of these methods and their dependence on other variables.  Finally we discuss future work and how best to integrate these techniques into a larger program for the next runs of the LHC.

\section{Jet-Veto and \pt Resummation Theory}
\label{theory} \setcounter{equation}{0} 

In this section, we briefly review the relevant resummations of large logarithms  in non-inclusive measurements that arise at higher orders in 
perturbation theory. In particular, we are interested in resummation of logarithms that arise in the presence of a jet-veto
or in the measurement of \pt distributions of $WW$, as well as the correlations between the two. 
In either case, the presence of large logarithms is a consequence of the presence of multiple scales in the problem. 
Besides the scale of the hard interaction $M$, non-inclusive measurements introduce additional scales, 
$\ptv$ for jet-veto measurement and $p_T^{WW}$for \pt distribution measurement, leading to logarithms of the form 
$\alpha_s^n \log^m(p_T/M)$ and $\alpha_s^n \log^m (\ptv/M)$ respectively at higher orders with $m \le 2n$.  

We now briefly describe resummation of large logarithms as implemented in this paper.   The resummation of logarithms from jet-vetos can be done directly in QCD~\cite{Banfi:2012yh}, but it naturally can also be expressed in 
soft-collinear effective theory (SCET)~\cite{Berger:2010xi}. For the \ww process we employ the SCET calculation as described in \cite{Jaiswal:2014yba}. 
The EFT is matched to the full theory of QCD at a hard scale $\mu_h \sim 2 m_W$
\footnote{Ref. \cite{Jaiswal:2014yba} employed the choice $\mu_h \sim M$ to minimize logarithms of the form $\log(\mu_h/M)$. Further, the default choice of the hard matching scale in \cite{Jaiswal:2014yba} was chosen to be $\mu_h^2 = - M^2$ to resum 
$\pi^2$ terms (see \cite{Jaiswal:2014cna} for further discussion). However, in order to facilitate comparison with \cite{Meade:2014fca}, the default matching scale in this paper is chosen to be $\mu_h = 2 m_W$.}.
Using the power counting parameter  $\lambda = \ptv/M$,  
 the matching coefficient is 
 renormalization-group (RG) evolved to a soft-scale, $\mu_f \sim \ptv$ characterizing the initial-state radiation (ISR). 
 The RG evolution of the matching coefficient resums large logarithms of the form $\log \lambda$. The factorized jet-veto 
 cross-section in SCET can be parametrically written as   
 \beq
\frac{\der \sigma(\ptv)}{\der M} \sim H(\mu)  Z_S(\mu, \nu, \bar{\nu}) B(\mu, \nu) \bar{B}(\mu, \bar{\nu}) 
\eeq
where, $M=M_{WW}$,  the {\it hard} function $H$ is the square of the matching coefficient, $B$ and $ \bar{B}$ are the collinear and 
anti-collinear {\it beam}-functions which describe ISR in the presence of jet-veto, and $Z_S$ is a renormalization 
constant for the product of beam functions, also referred to as {\it soft}-function in the literature. 
The beam functions have additional (rapidity) divergences which 
are not regulated by dimensional regularization and need additional regulators.  Associated with these additional regulators
are the renormalization scales $\nu$ and $\bar{\nu}$ as well as corresponding RG equations \cite{Jaiswal:2015nka}.  After implementing RG evolution in the $\mu$--$\nu$ space, the product of beam functions and soft-function in the factorized cross-section takes the form :
 \beq
Z_S(\mu, \nu, \bar{\nu}) B(\mu, \nu) \bar{B}(\mu, \bar{\nu}) = \left( \frac{\mu^2}{M^2} \right)^{g(\mu)} \hat{Z_S} \hat{B} \hat{\bar{B}} (\mu)
\label{eq:beam}   
\eeq
where the expressions for $g$,  $\hat{Z_S}$, $\hat{B}$ and $\hat{\bar{B}}$ as well as the procedure for estimating the scale uncertainties can be found in \cite{Jaiswal:2015nka}. 
The beam functions develop dependence on the jet-clustering parameter $R$ at $\co(\alpha_s^2)$,  which can lead to substantial scale uncertainties for small $R$ due to presence of $\log R$ terms which are not resummed in the current implementation.

To resum logarithims of the form $\log (p_T/M)$ we use the formalism of~\cite{Bozzi:2005wk} which was  implemented for the \ww channel LHC measurements in~\cite{Meade:2014fca}. The resummed partonic cross section takes the form 

\begin{equation}
\frac {d \hat \sigma_{ab}^{WW}}{dp_T^2} \left( p_T, M, \hat s, \alpha_s \left( \mu_R^2 \right), \mu_R^2, \mu_F^2 \right) = \frac {M^2} {\hat s} \int \frac{d^2 \mathbf b} {4\pi} e^{i \mathbf b \cdot p_T} \, \mathcal W_{ab}^{WW} \left( b, M, \hat s, \alpha_s \left( \mu_R^2 \right), \mu_R^2, \mu_F^2  \right).
\label{eq:bessel}
\end{equation} 
where $\mathcal W_{ab}^{WW} $ is the resummed cross section in impact parameter space ($b$-space).  The resummation is more easily performed after doing a further Mellin transformation which demonstrates the typical exponentiated structure
\begin{equation}
\mathcal W_{ab,N}^{WW} (b, M,\mu_F^2,\mu_R^2) = \mathcal H_{N}^{WW}(M,\mu_F^2,Q) \exp \left\{ \mathcal G_N (L,\mu_R^2,Q) \right\},
\end{equation}
where N is the moment of the Mellin transform with respect to $z=M/\hat{s}$, $\mathcal H_{N}^{WW}$ is the hard function, and $\mathcal G_N$ depends on physics at scales of $\sim p_T$.  We have introduced the new scale Q that accounts for the uncertainty associated with matching to the hard process and separating it into the various pieces, and finally $L=\log Q^2 b^2/b_0^2$ with $b_0$ a fixed constant of $\mathcal{O}(1)$.  In understanding the uncertainties associated with the resummed calculation we vary $\mu_F,\mu_R$ and $Q$, where $Q$ is expected to be similar to $M$ but below it.   NP effects can be systematically included to \pt resummation, however, for the distributions discussed in this paper the effects are small and the interested reader can find more details in~\cite{Meade:2014fca}.

 In both \cite{Jaiswal:2014yba}  and \cite{Meade:2014fca} the resummation was carried out to order NNLL+NLO which matched to NLO fixed order cross-section rather than NNLO, given that the full NNLO cross section was not yet available. With the calculation of the NNLO cross section for \ww production\cite{Gehrmann:2014fva}, it is possible to extend this analysis to NNLL+NNLO. This was performed for \pt resummation in \cite{Grazzini:2015wpa}. However work on similar next order analysis for jet veto resummation is still ongoing \cite{ongoing}. In order to compare resummations at the same order we use NNLL+NLO for both resummations in this paper.

The correlations between jet-veto and \pt resummation are most evident when one looks at large logarithms at {\it fixed order} 
in perturbation theory. To study the correlation, we focus on the leading-jet \pt which can be described by 
$\der \sigma (\ptv) / \der \ln \ptv$. At $\co(\alpha_s)$, leading-jet \pt is exactly balanced by the \pt of the $WW$ 
system and therefore, $\der \sigma (\ptv) / \der \ln \ptv = \der \sigma (p_T) / \der \ln p_T$. At $\co(\alpha_s^2)$, situation is complicated 
by the fact that more than one emission is allowed and the leading-jet \pt is no longer equal to the \pt of the $WW$ 
system. Nevertheless, given the similarity in structure of IR singularities, we expect correlations among the 
two observables. Indeed at $\co(\alpha_s^2)$, the logarithmic singularities in the difference 
$\der \sigma (\ptv) / \der \ln \ptv - \der \sigma (p_T)/ \der \ln p_T$ evaluated 
at $p_T = \ptv$ arise entirely from jet-clustering effects \cite{Banfi:2012jm}. Although this does not 
 constitute a rigorous proof, it lends credence to the \pt reweighting technique as a means of estimating jet-veto efficiency. 
 
 Finally, we comment on underlying events (UE) or soft-physics, which is known to effect non-inclusive observables, 
 such as $p_T$ distributions or jet-multiplicity.
Some sources of soft-physics can be captured perturbatively via resummation, however 
NP effects  such as hadronization (characterized by scale $\Lambda \lsim$ GeV), although not calculable in perturbation theory, appear as power suppressed terms $\co(\Lambda/\ptv)$ in SCET when the beam functions are operator product expanded on to  parton distribution functions (PDFs). Following \cite{Becher:2013iya}, we parametrize NP effects for the jet-veto calculation by substituting $g(\mu)$ in Eq \ref{eq:beam} with 
\beq
g(\mu) \rightarrow g(\mu) - \frac{1}{2} \frac{\Lambda}{ \ptv}.
\label{eq:NP} 
\eeq    

\section{Reweighting MC events and Applications}
\label{reweighting} \setcounter{equation}{0} 

 Each of the resummation methods outlined in the previous sections makes an accurate prediction for a unique differential variable. For transverse momentum resummation it is the transverse-momentum of the diboson, while for jet veto resummation, it is the cross section of the zero-jet bin. Both methods are more accurate for their corresponding differential observables than combining a fixed order calculation with a parton shower, however inherently they are inclusive with respect to other observables.  As a result it is impossible to get a fully differential cross section solely from either of these resummation schemes. Theoretically this is fine, but the most important question is how to compare to experimental results.   To do so would require the unfolding of experimental events to make a prediction for a theoretical observable.  This leaves the results susceptible to inherent biases in the original events used to simulate the results which are then inverted to define an unfolding for experimental results.  A much more straightforward procedure is simply to provide experiments with MC events that they can pass through their own detector simulations and compare directly to data.  This is impossible with just the results of the resummation calculations, however a theoretical solution that avoids unfolding data is to reweight monte carlo events.  Reweighting techniques have been used in multiple experiments, and have been used both for reweighting to theoretical calculations as well as reweighting distributions based on experimental data.   For the purposes of this paper reweighting simply amounts to the following. Given a particular differential direction denoted by $\xi$ predicted from resummation, the resummed distribution is binned and a reweighting function is defined by
 \begin{equation}
\text{F[}\xi\text{}]=\frac{\text{Resummed bin[}\xi\text{]}}{\text{MC bin[}\xi\text{]}}.
\end{equation}
 
 In this section we describe reweighting methods for each resummation calculation.  For \pt reweighting we employ a technique similar to that used in HqT~\cite{deFlorian:2011xf} and in~\cite{Meade:2014fca} where the underlying MC events are reweighted by the \pt of the diboson predicted from \pt resummation.  This was employed by CMS for the 8 TeV \ww measurement\cite{CMS8full} using the results of~\cite{Meade:2014fca} and good agreement was found for the cross section measurements.  ATLAS has not employed such a method, but it would be interesting to see given that ATLAS has consistently released the distribution of $p_T(\ell^+\ell^-+\mathrm{MET})$ which is the \pt of the diboson system up to the contribution to MET from the resolution of QCD objects. 
 
 For jet-veto reweighting it is a more subtle question of how to reweight events.  At its core, jet-veto resummation only gives one number, the 0-jet bin cross section.  There was an attempt in~\cite{Becher:2014aya} to construct an automated jet veto resummation procedure that reweights madgraph events at LO or NLO.  Unfortunately, this doesn't solve the problem of interfacing with experimental results as only the LO version produces events. In this case distributions determined by QCD corrections may be inherently incorrect, e.g. the \pt of the diboson system will have a pole at \pt=0.  More generally, given the predicted 0-jet bin and the overall inclusive cross section, it is always possible to construct a crude two-bin reweighting function simply based on whether or not there is a reconstructed MC jet above or below a jet veto scale which can be applied to events simulated at LO or NLO and interfaced with a parton shower.  The data-driven normalization applied to the WW background for Higgs studies, roughly corresponds to such a crude two-bin reweighting, and hence studying its effects on other differential shapes acquires importance. However, in predicting differential shapes, for instance for the \ww measurement (or \ww background to $H\rightarrow$\ww), in the fiducial cross section a reweighting function constructed this way would simply be an overall K-factor since all MC events in the 0-jet bin would be weighted the same as they would all pass the jet-veto. As a result, the predictions for the shape of the fiducial cross section will by definition only be as good as the underlying Monte Carlo prediction.  
 
\begin{figure}[ht]
\centering

\begin{subfigure}{\textwidth}
\centering
 \includegraphics[width=0.35\textwidth]{{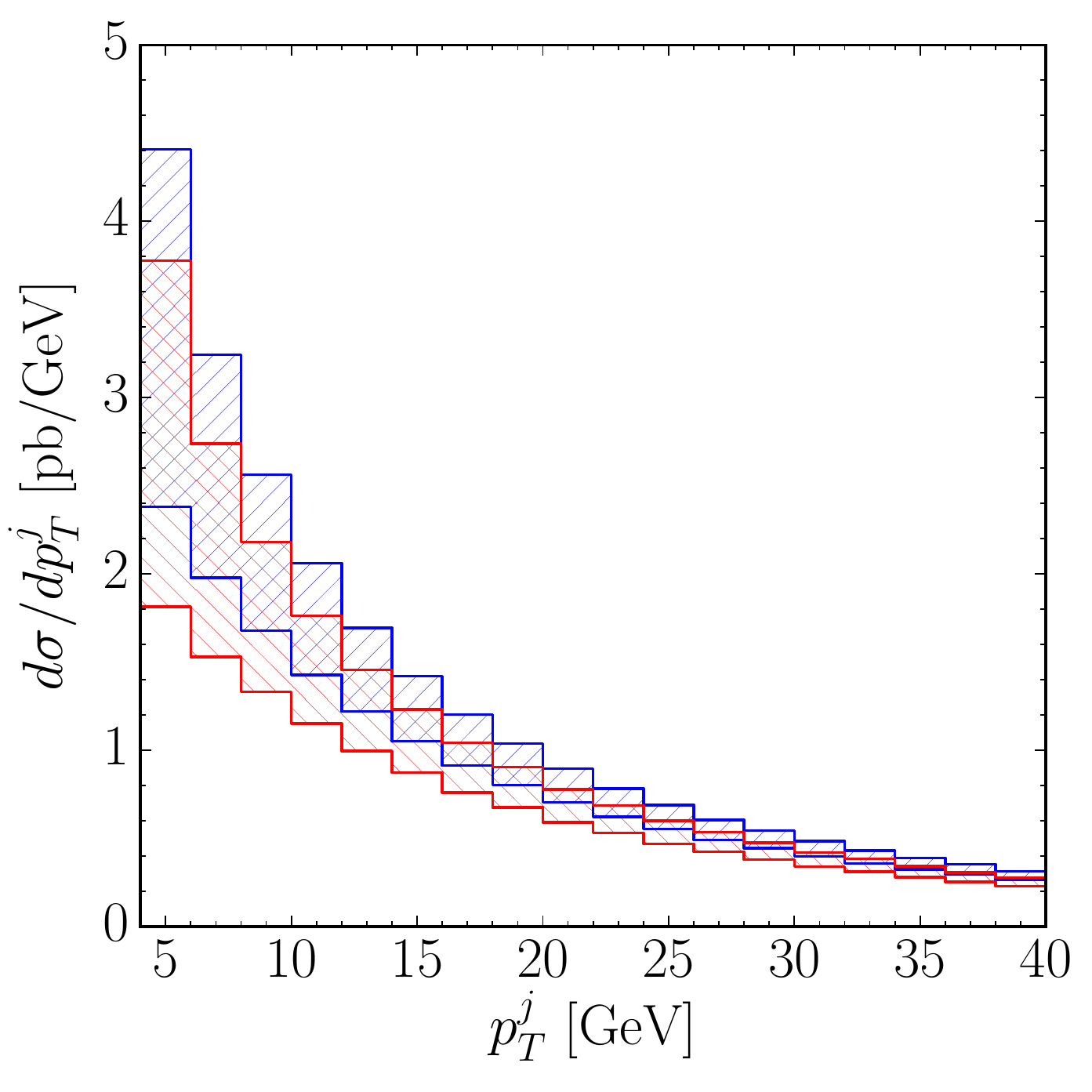}}  \includegraphics[width=0.35\textwidth]{{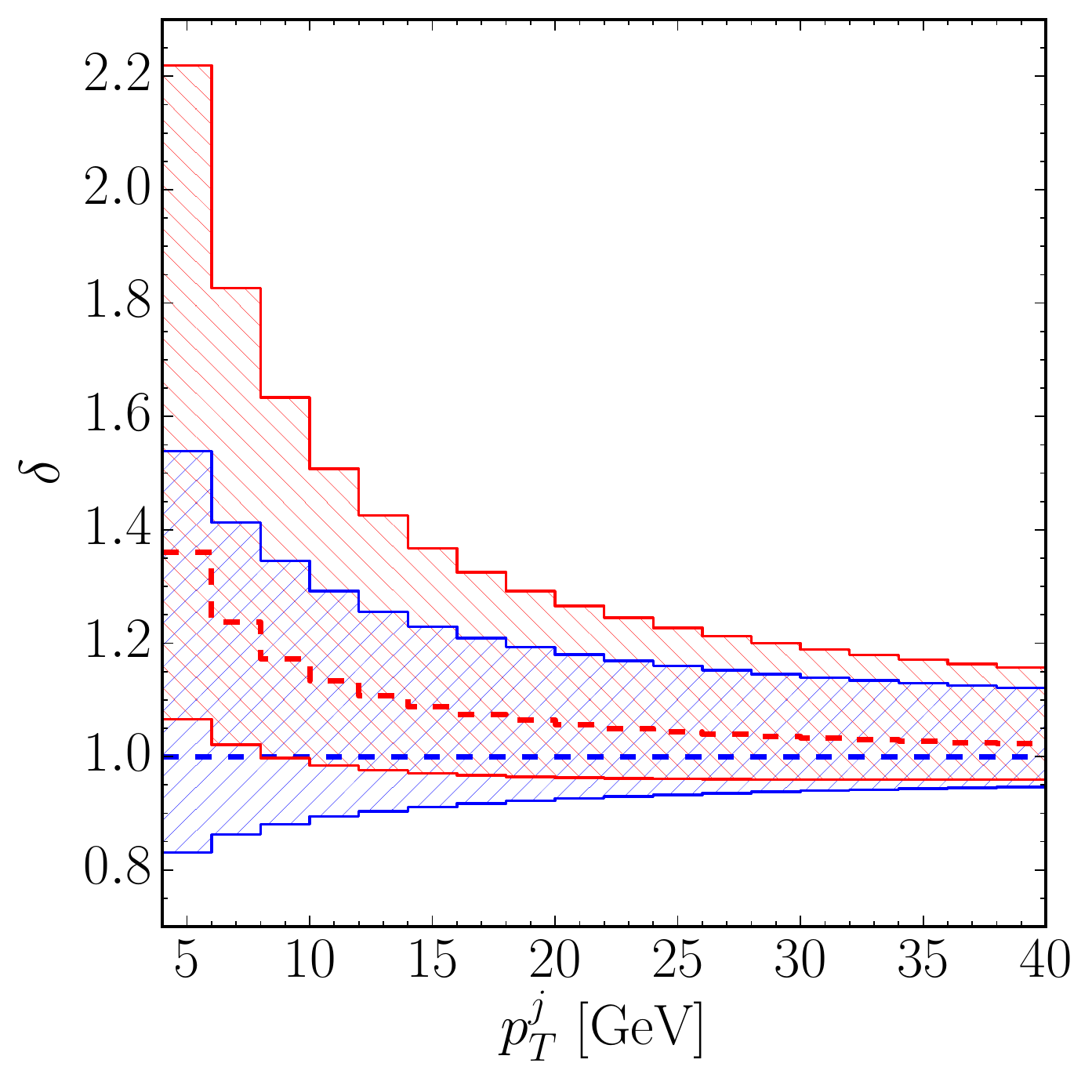}} 
\end{subfigure}
%
\begin{subfigure}{\textwidth}
\centering
 \includegraphics[width=0.4\textwidth]{{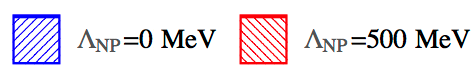}} 
\end{subfigure}
\caption{For $\sqrt{s}=8$ TeV and $R=0.4$ anti-k$_T$ jet algorithm, in the left hand panel $d\sigma/dp_T^j$ is plotted.  For the distribution shown in blue, errors come from scale variations without NP factors, in red $\Lambda_{NP}=500\,\mathrm{MeV}$ uncertainties are included.  In the right hand panel, the fractional uncertainty of $d\sigma/dp_T^j$ from scale variation relative to the central scale choice is shown with and without NP uncertainties.}
\label{relerrorptj}
\end{figure}

\begin{figure}[ht]
\centering

\centering
 \includegraphics[width=0.5\textwidth]{{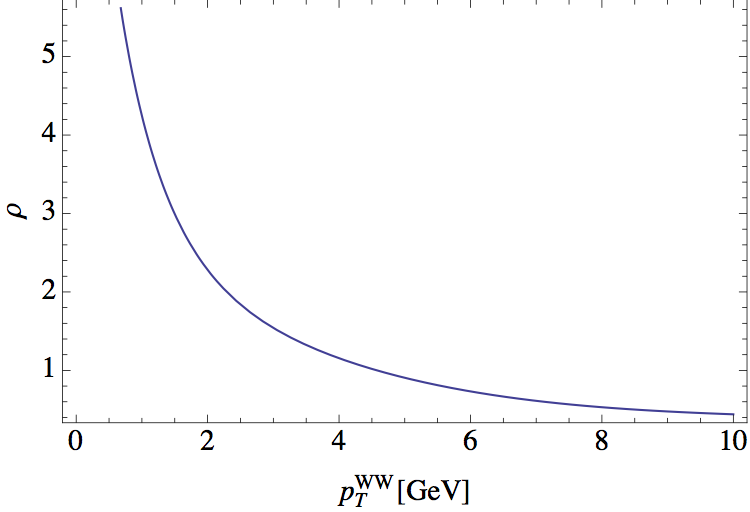}} 
\caption{correlation variable $\rho$ as a function of $p_T^{WW}$}
\label{corr}
\end{figure}

 A potentially more interesting possibility is to use the calculation of the jet-vetoed cross section from resummation to construct a continuous distribution for $\frac{d\sigma}{dp_T^j}$ where $p_T^j$ is the $p_T$ of the leading jet in the event.  However, depending on exactly how this is implemented there are issues with the size of the errors and the correlation to $p_T^{WW}$. Going to lower $p_T^j$ quickly leads to poor convergence of perturbation theory and eventually non-perturbative corrections take over. In Fig.~\ref{relerrorptj},  $\frac{d\sigma}{dp_T^j}$ is plotted as a function of $p_T^j$ by differentially binning the jet-veto 
 cross-section $\sigma(\ptv)$ (see Section~\ref{theory}) with respect to $\ptv$. To estimate the effects of NP corrections, we have also implemented $\Lambda = 500$ MeV in Eq~\ref{eq:NP}\footnote{In~\cite{Becher:2014aya} a similar uncertainty was estimated and they further attempted to quantify this effect by turning on/off hadronization in Pythia, resulting in a fit of $\Lambda = 240$ MeV.   Given the inherent uncertainty associated with how the many contributions to soft physics are taken into account in Pythia we use $\Lambda = 500$ MeV simply as an example as the ultimate uncertainty may be even larger.}. The relative scale uncertainty in $\frac{d\sigma}{dp_T^j}$, $\delta$ for each bin is estimated by the corresponding scale uncertainty in the jet-veto cross-section. The relative scale uncertainties normalized with respect to $\Lambda=0$ central values are also in Fig~\ref{relerrorptj}. Given the large errors at low $p_T^j$, reweighting at low $p_T^j$ is not advisable.   
Even if one is able to reduce scale uncertainties by going to high orders and further, devise methods to systematically quantify the NP corrections, there is still the problem of poor correlation between $p_T^j$ and $p_T^{WW}$ at low $p_T$. This is quantified in Figure~\ref{corr} where we plot the difference between $p_T^j$ and $p_T^{WW}$ as a percentage of $p_T^{WW}$ for a Powheg Monte Carlo WW sample showered with Pythia8,
\begin{equation}
\rho(p_T)=\frac{\langle \lvert p_T^j(p_T)-p_T \rvert \rangle}{p_T}
\end{equation}
where $p_T$ refers to $p_T^{WW}$.
For the above reasons we conclude that the naive two-bin reweighting method is the safest way forward if reweighting with jet-veto resummation has to be employed and we use only this method in the next section.

 Both \pt and jet-veto reweighting methods have their own advantages and disadvantages.  In \pt reweighting, all jet dependent effects are inherited completely from the underlying MC given that \pt resummation sums over all gluons and is fully inclusive.  This makes it impossible to estimate systematics on the jet-vetoed cross section from a purely theoretical viewpoint alone.  However, as we will show we find good agreement for the jet-veto efficiency between \pt reweighting and the jet-veto resummation calculation.   The jet-veto reweighting method gives the most theoretically under control calculation of the jet-veto efficiency, but all differential quantities including the \pt of the system will be essentially the same as for the MC as we show in the next section.  Therefore without joint resummation, we will show that \pt reweighting gives the best overall predictions of rate and shape in this channel. This is important since the ATLAS measurements of the \pt of the diboson system in the fiducial region disagree with the NLO+parton shower predictions, and jet-veto reweighting would predict the same distribution as the MC.

\section{Results and Comparison}
\label{results} \setcounter{equation}{0} 

\begin{figure}[ht]
\centering

\begin{subfigure}{\textwidth}
\centering
 \includegraphics[width=0.35\textwidth]{{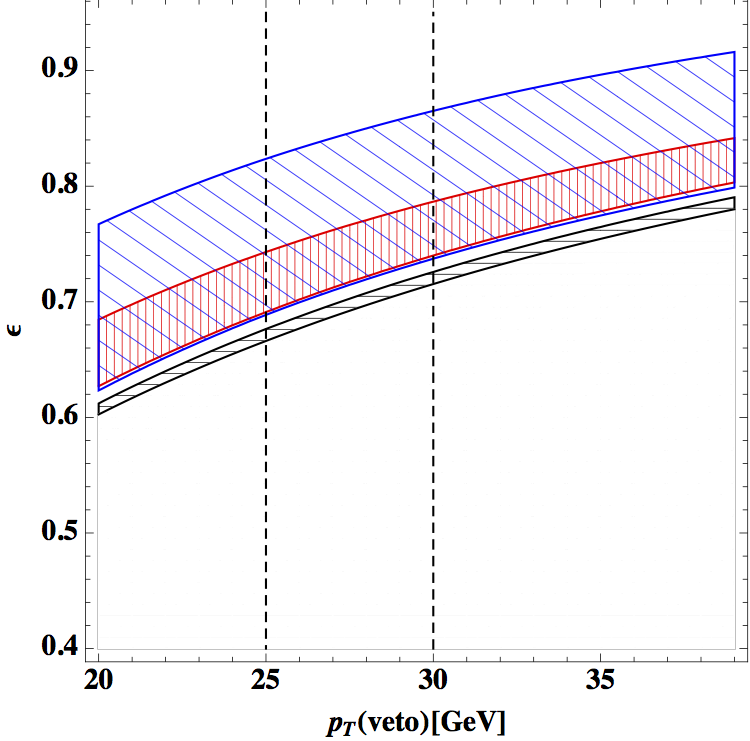}}  \includegraphics[width=0.35\textwidth]{{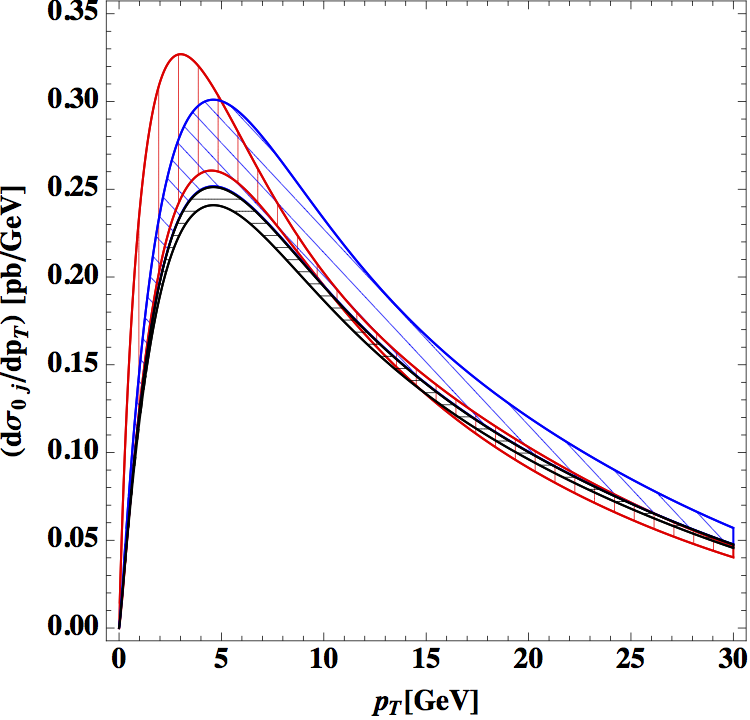}} 
\caption{8 TeV}
\end{subfigure}
\begin{subfigure}{\textwidth}
\centering

 \includegraphics[width=0.35\textwidth]{{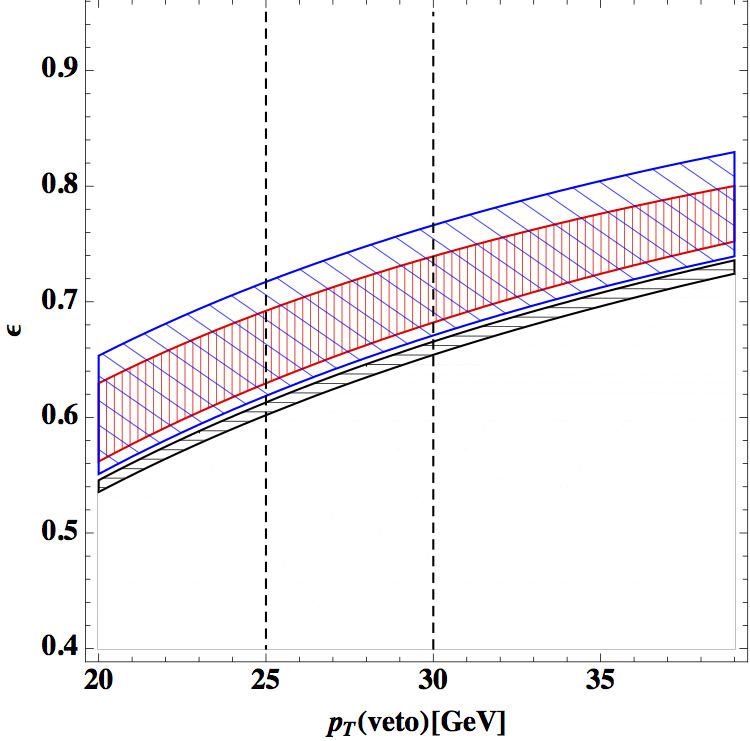}}  \includegraphics[width=0.35\textwidth]{{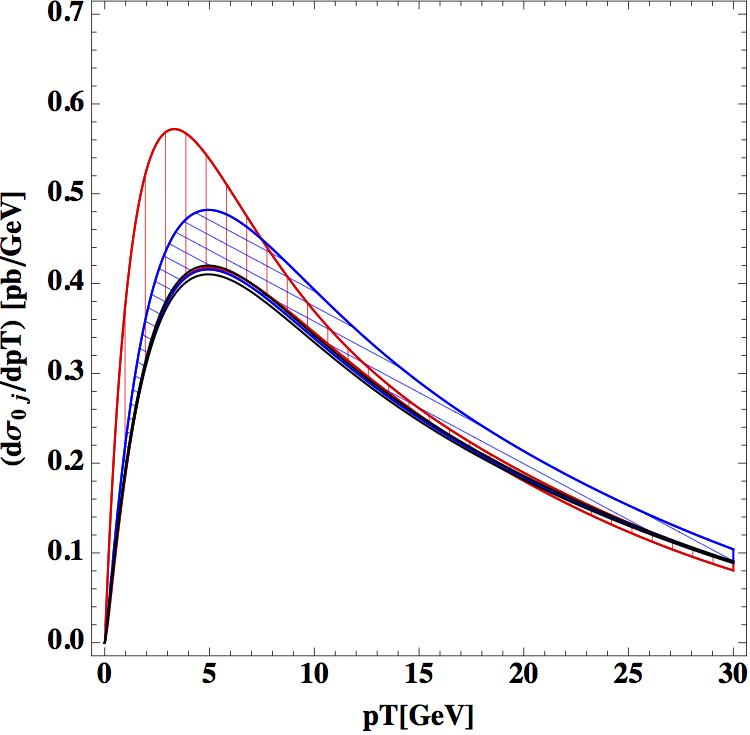}} 
\caption{13 TeV}

\end{subfigure}

%
\begin{subfigure}{\textwidth}
\centering
 \includegraphics[width=0.6\textwidth]{{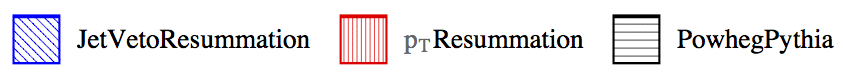}} 
\end{subfigure}
\caption{Comparison of jet-veto efficiency and $p_T^{WW}$ in the zero jet bin, from jet veto resummation and pT resummation for R=0.4 at 8 TeV (top) and 13 TeV (bottom).}
\label{jvvspt}
\end{figure}

The {\tt Powheg+Pythia} events for the process $ p p \rightarrow W^+ W^-$ at NLO are reweighted using two procedures, utilizing two different resummations,  as described in the previous section. These are then used to calculate jet-veto efficiency and the $p_T$ shape of the $WW$ system in the zero-jet bin. We have consistently used the {\tt MSTW2008nlo} PDF sets in both the resummation calculations and MC simulations. Earlier efforts at addressing these procedures separately Ref. \cite{Jaiswal:2014yba},\cite{Meade:2014fca} employed different scale choices. In this paper, we choose the same hard scale of $2 M_W$ to compare both procedures. We also turn off $\pi^2$ contributions that affect the overall cross-section.   However, there are still additional scale choices that must be made independently because of the different formalisms for the calculations, that we now list and we describe their impact on the uncertainty. For \pt resummation the renormalization (hard scale) and factorization scales are taken to be $\mu_R=\mu_F=2M_W$, and there is an additional resummation scale Q whose central scale is chosen to be  Q$=M_W$.  For jet-veto resummation, the central values for the hard scale and the factorization scale are chosen to be $2 M_W$ and $\ptv$ respectively. To capture scale variations, we vary the hard scale and factorization scale by a factor of 2 and $\frac{1}{2}$ around their respective central values and add the resulting errors in quadrature.  Additionally, there are scale uncertainties associated with rapidity divergences in SCET for which we follow the prescription outlined in \cite{Jaiswal:2015nka}. For $p_T$ resummation, based on the nature of the calculation we vary $\mu_F$ and $\mu_R$ in a correlated way by a factor of 2 around the central scale, and separately vary the resummation scale by the same factors and then add the resulting errors in quadrature. We have used the anti-k$_T$ jet algorithm for jet-clustering in {\tt Pythia} and the same algorithm is employed in the jet-veto resummation calculations. The analyses were performed both at 8 TeV as well as 13 TeV. These results are plotted in Figure~\ref{jvvspt} and the central values of the jet veto efficiencies relevant to ATLAS and CMS with their corresponding 1$\sigma$ errors are given in Table~\ref{table:tab1}. The left plots show jet-veto efficiency $\epsilon$, the fraction of events passing the jet veto, predictions directly from resummation (blue) as well as that obtained from  $p_T$ reweighting (red). This is to be compared with jet-veto efficiencies from {\tt Powheg} (black), which predicts a slighty lower efficiency. The two resummation methods however agree within 1$\sigma$ error bars at both 8 TeV and 13 TeV. This demonstrates that the existing procedure to use $p_T$ resummation to estimate the jet-veto cross-section is reliable. On the other hand, the $p_T$ shape of the \ww system in the zero-jet bin, shown on the right hand side of Figure~\ref{jvvspt}, as predicted by our jet-veto resummation reweighting retains the peak position of the underlying {\tt Pythia} sample while $p_T$ resummation predicts a softer $p_T$ shape\footnote{It should be noted that $p_T$($\ell \ell$ + $\MET$) is the actual measurable quantity that corresponds to the $p_T^{WW}$ shape and this suffers from significant $\MET$ smearing.}.

\begin{center}
\begin{table}[ht]
\begin{centering}
  \begin{tabular}{ | l || c  | c |   }
    \hline
      & \multicolumn{2}{ c| }{Jet Veto Efficiency} \\ \cline{1-3}

    $\ptv$  &  25 GeV & 30 GeV \\ \hline \hline
    8 TeV $p_T$  resummation  & ${0.71}^{+0.03}_{-0.02}$ & ${0.76}^{+0.03}_{-0.02}$ \\ \hline
    8 TeV Jet-veto resummation  & ${0.73}^{+0.09}_{-0.05}$ &${0.78}^{+0.09}_{-0.05}$ \\ \hline \hline
    13 TeV $p_T$  resummation  & ${0.66}^{+0.04}_{-0.03}$ &${0.71}^{+0.03}_{-0.03}$ \\ \hline
    13 TeV Jet-veto resummation  & ${0.65}^{+0.07}_{-0.03}$ &${0.70}^{+0.07}_{-0.03}$ \\ \hline
    
\end{tabular}
\caption{Jet Veto Efficiency at 8 and 13 TeV for $R=0.4$ }
\label{table:tab1}
\end{centering}
\end{table}
\end{center}

\subsection{Jet Definitions and other QCD effects}

\begin{figure}[ht]
\centering

\begin{subfigure}{\textwidth}
\centering
 \includegraphics[width=0.3\textwidth]{{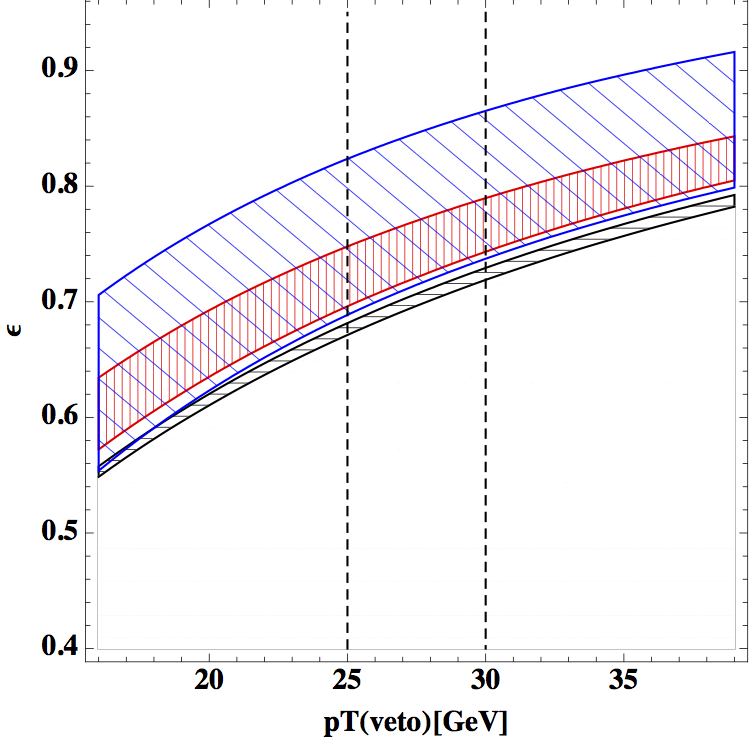}}  \includegraphics[width=0.3\textwidth]{{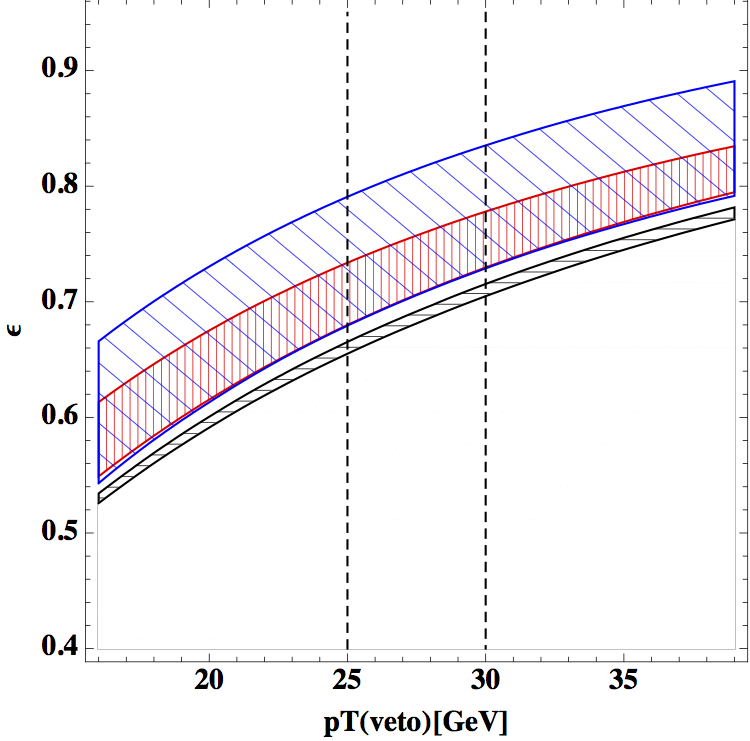}} \includegraphics[width=0.3\textwidth]{{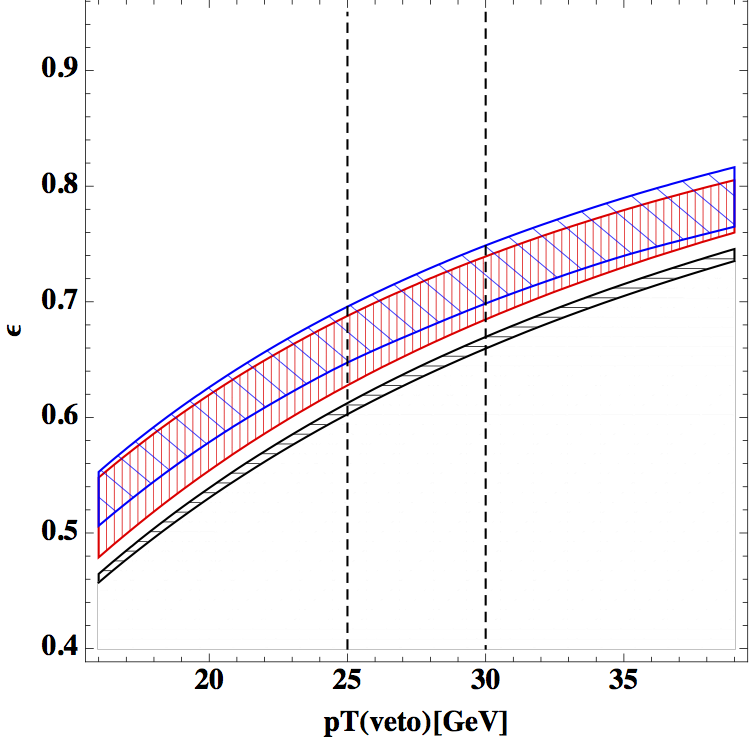}} 
\end{subfigure}

%
\begin{subfigure}{\textwidth}
\centering
 \includegraphics[width=0.6\textwidth]{{key.png}} 
\end{subfigure}
\caption{Comparison of jet veto efficiencies for 8 TeV for $R$=0.4, 0.5 and 1.}
\label{Rvariation}
\end{figure}

The jet-veto resummation calculation contains dependence on the jet-algorithm through $R$-dependent terms, which for small $R $ are dominated by $\log R$ terms arising from collinear splitting. As shown in  Figure~\ref{Rvariation}, the agreement between jet veto and $p_T$ resummation is better for large $R$. This is expected since, for larger $R$, more of the hadronic activity is captured as a single jet and hence the correlation between the leading jet momentum and \ww momentum is higher. While moving to $R \sim 1$ reduces the scale uncertainty in jet-veto resummation, due to better control of $\log R$ terms in perturbation theory, MPI effects can be quite large for large $R$ making the choice $R\sim 1$ far from ideal as we discuss below.

\begin{figure}[ht]
\centering

\begin{subfigure}{\textwidth}
\centering
 \includegraphics[width=0.3\textwidth]{{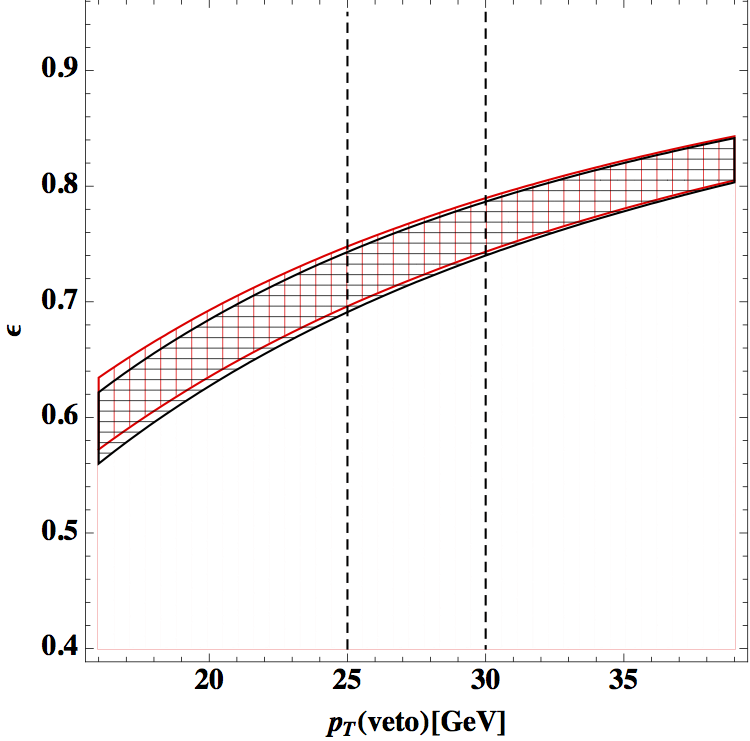}}  \includegraphics[width=0.3\textwidth]{{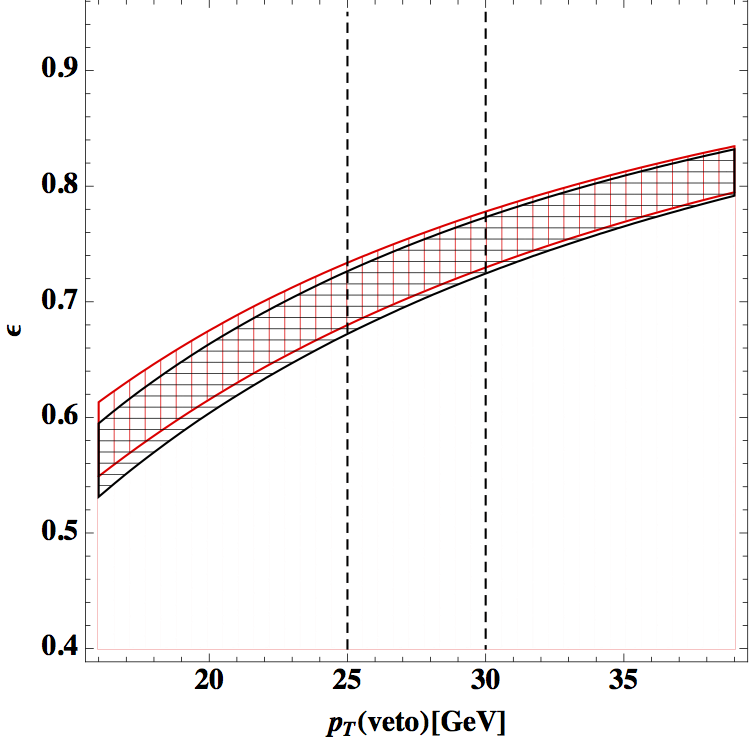}} \includegraphics[width=0.3\textwidth]{{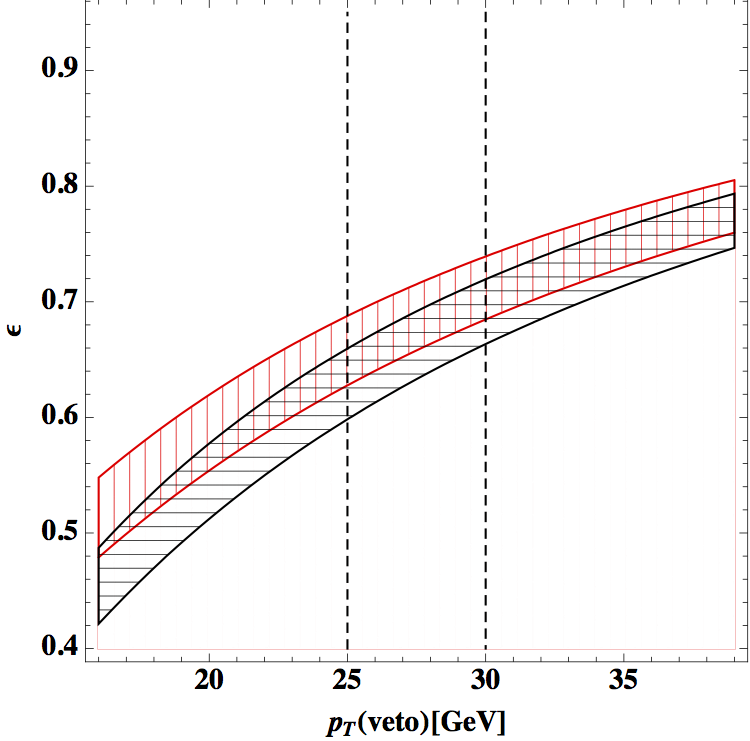}} 
\end{subfigure}

%
\begin{subfigure}{\textwidth}
\centering
 \includegraphics[width=0.3\textwidth]{{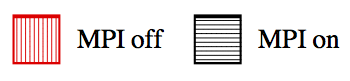}} 
\end{subfigure}
\caption{Comparison of jet veto efficiencies for 8 TeV using pT reweighting method with MPI off vs on for R =0.4, 0.5 and~1.}
\label{MPIoffvson}
\end{figure}

While the $p_T$ reweighting technique, which is inclusive in hadronic activity, is robust against MPI and NP effects such as hadronization, the same is not true for jet-$p_T$ reweighting technique. The $p_T$ distribution of a colorless-final state, such as \ww, will be practically unaffected by MPI since the soft-radiation associated with MPI is mostly isotropic. On the other hand, observables involving jets in the final state, such as jet-veto efficiencies, will be strongly effected by MPI. Jets with large $R$ contain more soft-radiation and therefore are prone to larger MPI effects.  We confirm this effect, that while turning off MPI does not affect the $p_T^{WW}$ shape, it does have an effect on jet-veto efficiency especially for large $R$ as shown in Fig  \ref{MPIoffvson}. To avoid such contamination, we recommend that LHC experiments continue to choose $R \sim 0.4$ in the \ww channel thereby minimizing dependence on MPI modeling. Further, for the case of jet-vetoes, MPI effects will be more pronounced for small $\ptv$ where the soft-radiation starts competing with the {\it true partonic} jet $p_T$. In order to minimize the impact of NP effects on jet $p_T$ reweighting, a large underflow bin in jet-$p_T$ was chosen. With this choice of underflow bin, the $p_T$ shapes are fairly independent of whether MPI and hadronization effects in {\tt Pythia} are included or not, as long as the jet radius parameter $R$ is not too large\footnote{By default, we have turned on MPI and hadronization effects in {\tt Pythia}.}.


\section{Discussion}
\label{s.discussion} \setcounter{equation}{0} 

In this paper we have explored the agreement of different resummation procedures for predicting both the \ww fiducial cross section with a jet veto as well as the differential cross section with respect to the diboson \pt. Prior to this paper there was some confusion surrounding the different resummation methods and whether they led to different conclusions~\cite{Becher:2014aya}.  The reflection of this is most clearly represented in the most recent ATLAS and CMS measurements, where CMS chose to use the resummation improved theory predictions and found agreement with the SM while ATLAS did not and has a discrepancy which persists.   Nevertheless, in this paper we have shown that when comparing jet veto efficiencies directly, without modifying the inclusive cross section, the methods of \cite{Meade:2014fca} and \cite{Jaiswal:2014yba} agree very well.  In particular, we have identified that the predictions at both Run 1 and Run 2 of the LHC should agree within perturbative uncertainties for each method, as well as with any reasonable choices of jet definitions or variation of non-perturbative sources of error.  Therefore regardless of the method, ATLAS and CMS should use some form of resummation when comparing to theoretical predictions to describe the \ww fiducial cross section with a jet veto.

In addition to the results for the jet vetoed cross section, we have also implemented a reweighting procedure based upon jet-veto resummation so that differential predictions can be compared between the two methods.  For instance since \pt resummation by construction best predicts the \pt distribution of the diboson system, it's useful to  compare the predictions from the jet-veto reweighted method as shown in Figure~\ref{jvvspt}.  The increase at low \pt compared to {\tt Powheg}-{\tt Pythia} is noticeable for both methods which would for instance bring the ATLAS experimental data for $p_T (\ell^+\ell^-+\MET)$ into better agreement, however, there are still noticeable differences.  In particular, \pt resummation predicts a \pt distribution that peaks at lower \pt than the jet-veto reweighting procedure and the MC prediction.  To compare these methods further there are both experimental and theoretical opportunities and challenges.  In measuring the \pt of the system of the system experimentally for the \ww channel there inherently will be smearing due to the $\MET$ resolution.  

\begin{figure}[ht]
\centering

\begin{subfigure}{0.5\textwidth}
 \includegraphics[width=\textwidth]{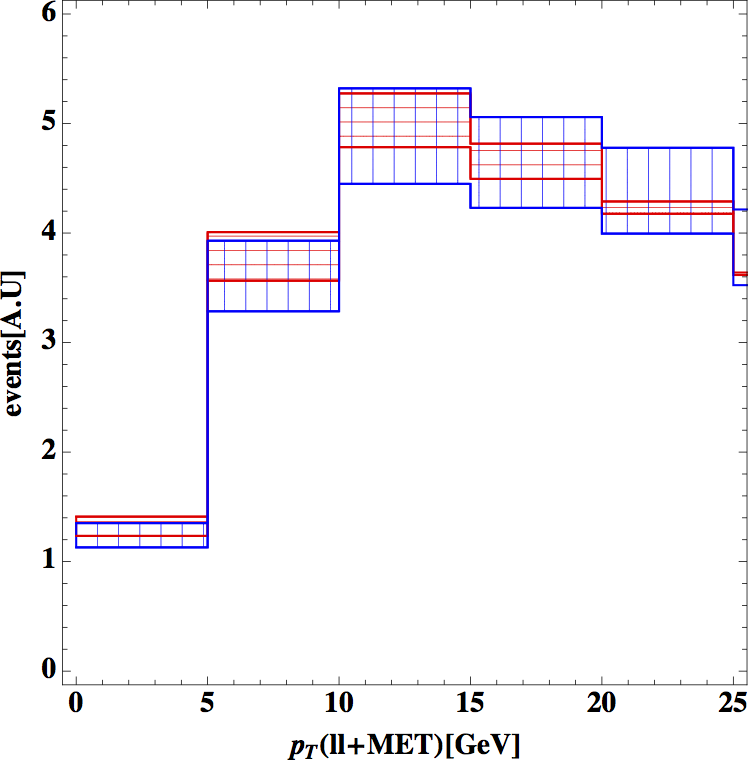} 

\end{subfigure}
\begin{subfigure}{0.5\textwidth}
\includegraphics[width=\textwidth]{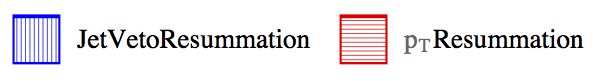}
\end{subfigure}
\caption{The  $p_T (\ell^+\ell^-+\MET)$ distribution after a parametrized smearing of $\MET$.  }
\label{metsmear}
\end{figure}

We demonstrate this in Figure~\ref{metsmear} by applying a $\MET$ smearing\footnote{We find smearing parameters for $\MET$ parallel and perpendicular to $p_T^{WW}$by fitting to the \pt$(\ell^+ \ell^-+\MET)$ plot in \cite{atlas8ww}.} to the predictions for the $p_T (\ell^+\ell^-+\MET)$ distribution from both \pt and jet-veto resummation.  Unfortunately even though at truth level there are theoretically different distributions, the difference are washed out in a channel such as \ww.  Nevertheless, it is important to note that the resummation calculations reviewed in this paper are essentially universal for all diboson processes.  Therefore, with the high luminosity run of the LHC it should be possible to disentangle these effects and ultimately provide a robust test of whether QCD can successfully describe these important proves of the EW structure of the SM.   In addition, it would be useful to find new variables in MET dominated channels that remove the sensitivity to the $\MET$ uncertainty and allow for further quantitative tests of QCD predicted by resummation.

Theoretically without relying on reweighting, the only way to advance further in the directions we have explored is to go to a joint resummation formalism such as in~\cite{Procura:2014cba}.  While this program is and should be carried out, much information can be gained by comparing across channels with and without jet-vetoes to better choose scales and NP factors in \pt resummation.  At this point, with the lessons learned from comparing individual resummation methods we recommend that experimentalists implement \pt resummation reweighting for all diboson channels.   This should be extended not only to the SM measurements but to background processes as well, for instance for $h\rightarrow W^+W^-$.  A final theoretical error in jet-vetoed processes can be formed from jet-vetoed cross sections, but in practice at this point \pt resummation is most useful to reweight events and better to directly compare with experimental data.  

As we have demonstrated, we have entered into a qualitatively new era at the LHC where we now have an example of the need for NNLL resummation in addition to NNLO fixed order calculations to describe the data.   To go further will require both theoretical and experimental efforts.  It is important that a program be developed that cuts across various SM channels and provides a comprehensive test of how well the SM describes LHC data.  While we have shown that higher order QCD corrections can ameliorate the most discrepant SM measurement from Run 1 of the LHC, it is important to note that none of these effects were included in other SM channels.  In particular, the stunning agreement with the SM in almost all channels compared to NLO MC results and inclusive cross sections should now be interpreted as a systematic discrepancy in almost all diboson channels other than $W^+W^-$.  It is important to investigate this further, and we hope with a concerted theoretical and experimental effort at Run 2, we will see whether the SM triumphs at the EW scale or we will have our first hints of new physics emerging. 

\subsection*{Acknowledgements}

We would like to thank Thomas Becher, Valentin Hirschi, Rafael Lopes de Sa and Mao Zeng for useful discussions. The work of P.M. was supported in part by NSF CAREER Award NSF-PHY-1056833.  The work of H.R  was supported in part by NSF grant PHY-1316617.
 

\end{document}